**Equilibration and Filtering of Quantum Hall Edge States in Few-Layer Black Phosphorus**

Jiawei Yang[1], Kangyu Wang[1], Shi Che[1], Zachary J. Tuchfeld[1], Kenji Watanabe[2], Takashi Taniguchi[3], Dmitry Shcherbakov[1], Seongphil Moon[4], Dmitry Smirnov[4], Ruoyu Chen[1], Marc Bockrath[1], Chun Ning Lau[1*]

[1] Department of Physics, Ohio State University, Columbus, OH 43220
[2] Research Center for Functional Materials, National Institute for Materials Science, 1-1 Namiki, Tsukuba 305-0044, Japan
[3] International Center for Materials Nanoarchitectonics, National Institute for Materials Science, 1-1 Namiki, Tsukuba 305-0044, Japan
[4] National High Magnetic Field Laboratory, Tallahassee, FL 32310

**Abstract**

**We realize *p-p'-p* junctions in few-layer black phosphorus (BP) devices, and use magneto-transport measurements to study the equilibration and transmission of edge states at the interfaces of regions with different charge densities. We observe both full equilibration, where all edge channels equilibrate and are equally partitioned at the interfaces, and partial equilibration, where only equilibration only takes place among modes of the same spin polarization. Furthermore, the inner *p'*-region with low-doping level in the junction can function as a filter for highly doped *p*-regions which demonstrates gate-tunable transmission of edge channels.**

In a magnetic field, electron orbits in a two-dimensional electron system coalesce into Landau levels, giving rise to chiral quantum Hall edge (QH) states that propagate without dissipation. These edge states provide an attractive platform for exploring strong electron interactions, one-dimensional transport, interferometry and resistance standard[1-4]. More recently, manipulation of these edge states has been proposed as a route for topological quantum computation[5-7].

In the past decade, atomically thin two-dimensional (2D) semiconductors provide alluring platforms for control and manipulation QH edge states, with advantages such as tunability of band gap, effective mass, and charge density[8-12]. In particular, atomically thin black phosphorus has been intensively studied due to its high electron mobility and a direct band gap that is tunable by electric field, layer and strain, with wide range of potential electronic, optoelectronic and thermoelectric applications[13-24]. With recent improvement in materials and device preparation, we have recently achieved field effect mobility of up to 50,000 $cm^2/Vs$ and 4,000 $cm^2/Vs$ in BP field effect transistor devices at low temperature and room temperature, respectively, and observed integer and fractional QH states under high magnetic field[25]. Yet, the interaction and equilibration of QH edge states in few-layer BP devices have not been studied.

Here we report magnetotransport measurements of few-layer BP devices with local gates, thus realizing *p-p'-p* junctions with tunable densities in different regions. In the QH regime, the equilibration of edge states at the interface of regions with different doping levels gives rise to integer and fractional plateaus in two-terminal conductance and four-terminal longitudinal resistance values. In devices with moderate mobility, the plateau values are well accounted for by a model that assumes full equilibration among all edge states. Surprisingly, in a device with higher mobility, the plateaus values suggest separate, selective equilibration among channels with different spin polarization. When the outer *p*-regions are highly doped, the junction can also serve as a gate-tunable quantum filter that can shut off the transmission channels one by one. As the first observation of partial equilibration and selective transmission of edge states in a 2D semiconductor, our work underscores the potential afforded by 2D materials for exploring and harnessing novel phenomena and applications.

---

[*] Email: jeanielau1@gmail.com

We construct the black phosphorus devices by using polypropylene carbon (PPC) films to sequentially pick up hexagonal boron nitride(hBN), BP and hBN flakes. The completed hBN/BP/hBN stacks are released onto $Si/SiO_2$ wafers with pre-patterned local metal gates that are either continuous or split with a ~150 nm-wide gap. All exfoliation and transfer steps are completed inside a glove box, with moisture and oxygen concentration lower than 0.1 ppm. The stack is patterned into Hall bar geometry by $SF_6$ plasma reactive-ion etching (RIE), with metal gate located across the center of the Hall bar. Contacts to BP are made by patterning with electron beam lithography and RIE that etches only the top hBN layer. A schematic of device is shown in Fig.1a. The metal gate controls the charge density of the central region of the device, while the Si back gate controls that of the two outer regions. The devices are characterized at temperature T=0.3K using standard ac lock-in techniques.

Fig. 1b displays longitudinal resistance $R_{xx}$ of device D1, which has quantum mobility ~1500 $cm^2$/Vs, as a function of voltages applied to the Si back gate $V_{bg}$ and the metal gate $V_{mg}$. When either of the gate voltages is close to 0, the device is undoped and insulating; when the gate voltages decreases, the device becomes highly hole- or *p*-doped as its resistance decreases to a few kΩ. Fig. 1c display representative line trace from the color plot -- $R_{xx}(V_{bg})$ at $V_{mg}$=-1V is plotted as the red curve against the bottom axis, whereas $R_{xx}(V_{mg})$ at $V_{bg}$=-40V is plotted as blue curve against the top axis, . The inset of Fig. 1c is an optical image of a typical device. Thus the device acts as a tunable *p-p'-p* junction, where resistance of the central and the flanking regions can be separately tuned through several orders of magnitude, from undoped and insulating to highly hole-doped and conductive. At high magnetic fields, well-defined QH states are resolved[9, 13, 26, 27]. Fig. 1d displays $R_{xy}(V_{bg})$ of the side regions at $V_{mg}$=-6V and B=30T, where well-quantized resistance plateaus at filling factors *ν*=-1, -2, -3 and -4 are observed. We note that even though BP is anisotropic, no anisotropy is observed once it enters the QH regime[28].

To examine the equilibration of QH edge states, we measure D1's two-terminal conductance G at B=30T as $V_{bg}$ and $V_{mg}$ vary. Fig. 2a displays the resultant data, where the filling factors of the metal-gated region $ν_{mg}$ and of the Si-gated regions $ν_{bg}$ are labeled on the right and top axes, respectively. In the 2D color plot, the data appear as a plaid of colored squares, indicating that the regions are only controlled by a single gate, as expected. The conductance values depend on the filling factors of both regions. We first focus on the regime where $|ν_{mg}| > |ν_{bg}|$, and the central region acts as a scatterer for the incident edge states. In the simplest and most common scenario, charge carriers in the central region are equally partitioned among all modes present, regardless of their origination or quantum numbers such as spin, valley and Landau level indices[29-35]. The resultant conductance is smaller than either filling factor, and is often quantized at a fractional value of the conductance quantum $e^2/h$, where *e* is electron charge and *h* the Planck constant. Thus, within the model of full edge state equilibration, the two-terminal conductance of a p-p'-p junction is given by[30]

$$G \text{ (in units of } \frac{e^2}{h}) = \begin{cases} |\nu_{mg}| &, |\nu_{bg}| \geq |\nu_{mg}| \\ \frac{|\nu_{mg}||\nu_{bg}|}{2|\nu_{mg}|-|\nu_{bg}|} &, |\nu_{bg}| \leq |\nu_{mg}| \end{cases} \quad (1)$$

(the case of $|v_{bg}| \geq |v_{mg}|$ will be discussed at the end of the Letter).

Fig. 2b illustrates $G(v_{bg}, v_{mg})$ calculated using Eq. (1), where the expected plateau values are displayed as numbers within each square. The experimental data of $G(v_{bg})$ at $v_{mg}$=-2 and -4 are shown in Fig. 2c, and corresponding theoretical curves are shown in as dotted lines. Similar plots for $G(v_{mg})$ at constant $v_{bg}$ are displayed in Fig. 2d. The data are in excellent agreement with that expected from Eq. (1), thus suggesting full equilibration of QH edge states in this few-layer BP device. Some features in the data that are not captured by the equation, such as the small deviation of the plateau values and the resistance dips and finite transition width between the plateaus, may be attributed to impurity scattering and disorder-induced Landau level broadening[36, 37].

We now turn to a different device (D2) with higher mobility, ~2500 cm$^2$/Vs. It differs from device D1 in that it has a split (instead of continuous) metal gate. Nevertheless, this device also functions as a *p-p'-p* junction, since the narrow gap between the split gates (~150 nm) is considerably smaller than widths of the depletion (compressible) stripes[38], which is $l \sim \frac{V_g \epsilon}{4\pi^2 ne} \sim$ 250 nm, for typical operating parameters. This behavior is also verified in the regime of $|v_{mg}| < |v_{bg}|$, where both devices display exactly the same behavior (see Fig. 5 and associated discussion).

Fig. 3a plots the device's two-terminal conductance in unit of $e^2/h$ a function of $V_{mg}$ and $V_{bg}$, where the dashed lines denote partition of regions of different filling factors. The corresponding map calculated from Eq. (1) is shown in the left panel Fig. 3b. We first focus on regions with $v_{bg}$=-1, i.e. the vertical light blue stripe that is centered at $V_{bg}$~-9V. As shown by the line cut in Fig. 3c, $G$=1 for $v_{mg}$=-1, which is expected, as the entire device is tuned into the $v$=-1 QH state with a single chiral edge channel. However, the $G$=1 plateau persists to $v_{mg}$=-2, contradicting Eq. (1) that predicts $G$=2/3 (blue dotted line in Fig. 3c); at $v_{mg}$=-3, data indicate $G$=0.7 whereas Eq. (1) predicts 0.6. Such deviation between data and Eq. (1) is also observed for the region ($v_{bg}$=-2, $v_{mg}$=-3) (Fig. 3d), where the data and calculations based on Eq. (1) are shown as red solid and blue dotted lines, respectively.

The disagreement between the data and Eq. (1) prompts us to re-evaluate the assumption of full equilibration. In high mobility devices, QH states with different quantum numbers may not fully equilibrate, due to finite spatial separations and/or the absence of scatterers that can flip the quantum number. For instance, previous works in graphene reported edge states that only equilibrate with the same spin polarization[39] or Landau level (orbital) index[35]. In our case, the deviation from Eq. (1) is observed within the lowest Landau level. Thus we explore the scenario of spin-selective partial equilibration, i.e. equilibration is only achieved among edge states with the same spin, and absent among ones with different spins. In this scenario, $G = G_\uparrow + G_\downarrow$, where $G_\uparrow$ and $G_\downarrow$ are obtained by applying Eq. (1) to spin-up and spin-down edge states, respectively. For instance, for the case $v_{bg}$=-1 and $v_{mg}$=-2, the spin-up and spin-down edge states do not interact in the central region; hence, the single spin-up state is fully transmitted, while the single spin-down channel is localized within the central region and does not contribute to the conductance. This results in a total conductance of 1 $e^2/h$, in agreement with data. Similarly, for the case $v_{bg}$=-2 and $v_{mg}$=-3, the spin-up (spin-down) channel consists of 1 (1) and 2 (1) edge states in the outer and central regions, respectively, giving rise to $G_\uparrow$=2/3 and $G_\downarrow$=1, and a total conductance of 5/3 $e^2/h$ (Fig. 3d inset).

Using this model, the calculated $G$ values are shown in the right panel of Fig. 3b, and as red dotted lines in Fig. 3c-d. They are in much better agreement with the data than the full equilibration model, thus establishing that equilibration is only achieved among each spin species.

At the first glance, the different equilibration process in the two devices might be surprising, considering that the relatively modest increase in the magnitude of mobility (albeit 70% increase). However, mobility characterizes overall (and mostly bulk) scattering in the device, equilibration of the edge states depends on scatterers *close to the sample edges* and the *interfaces at the p-p' junctions*. Thus mobility is a relevant, albeit incomplete metric, for the equilibration process here.

Validity of this model is further verified by four-terminal resistance measurements of D2, where the metal-gated region is located between the two voltage probes. The longitudinal resistance is given by

$$R_{xx} = \frac{h}{e^2}\left|\frac{1}{|v_{mg}|} - \frac{1}{|v_{bg}|}\right| \qquad (2)$$

for the full equilibration model, and

$$R_{xx} = \frac{r}{T}\frac{1}{|v_{bg}|}\frac{h}{e^2} \qquad (3)$$

for partial equilibration, where $T=\frac{G_\uparrow+G_\downarrow}{|v_{bg}|}$ is the overall transmission coefficient for the $|v_{bg}|$ incoming channels, and $r=1-T$ is the reflection coefficient. Fig. 4a displays the $R_{xx}(v_{bg}, v_{mg})$ map calculated according to Eq. (3), and Fig. 4b the experimental data. Fig. 4c-d plots line traces $R_{xx}(v_{mg})$ at $v_{bg}$=-1 and -2, respectively, as well as expected values calculated from Eq. (2) (blue dashed lines) and Eq. (3) (red dotted lines). The data is most consistent with the partial equilibration model, again confirming spin-based equilibration.

Lastly, we investigate the regime with relatively large $|v_{bg}|$ so that $|v_{mg}| < |v_{bg}|$, and the central region hosts a smaller number of edge states than the outer regions. Here only $|v_{mg}|$ edge states are fully transmitted from source to drain, whereas the rest are reflected. Therefore the conductance is given by the upper equation of Eq. (1), and expected values are shown in Fig. 2b and 3b. Fig. 5a-b plots $G(v_{mg})$ at $v_{bg}$=-4 and -5 for device D1 and D2, respectively. In both graphs, as $|v_{mg}|$ is lowered, $G$ decreases stepwise from $|v_{mg}| = |v_{bg}|$ to 0, indicating that the edge channels are filtered out one by one, until conductance is completely turned off. Thus, in this regime, the *p-p'-p* junction serves as a "filter" for the edge channels, similar to the pinch-off of edge states by quantum point contacts in GaAs-based heterostructures. Such gate-tunable filters of edge states can be used to construct quantum computation platforms based on interferometry of edge states.

In summary, using dual-gated few-layer BP devices, we demonstrate that they function as *p-p'-p* junctions with tunable charge densities in each regions. In the QH regime, when the central region hosts a larger number of edge states than the outer regions, the circulating edge states therein couple modes that have different electrochemical potentials. We observe both full equilibration, where all modes equilibrate and are partitioned equally at the *p-p'* interfaces, and partial equilibration, where modes only equilibrate among the same spin polarization. When the central region hosts fewer edge states than the outer region, the junction behaves as a filter so as to enable stepwise, gate-tunable transmission of integer quantum Hall edge channels. These results underscore the potential of BP and 2D semiconductors as a versatile platform for understanding and manipulating QH physics and other topological edge states.

**Acknowledgement**
This work is supported by NSF/ECCS 1509958 and NSF/DMR 1807928. A portion of this work was performed at the National High Magnetic Field Laboratory, which is supported by National Science Foundation Cooperative Agreement No. DMR-1157490 and the State of Florida. K.W. and T.T. acknowledge support from the Elemental Strategy Initiative conducted by the MEXT,

Japan ,Grant Number JPMXP0112101001,  JSPS KAKENHI Grant Number JP20H00354 and the CREST (JPMJCR15F3), JST.

Fig. 1. (a). Schematics of few-layer BP devices. (b). Longitudinal resistance in unit of kΩ $R_{xx}$ vs. voltages applied to metal gate $V_{mg}$ and Si back gate $V_{bg}$. (c). Line traces $R_{xx}(V_{bg})$ at $V_{mg}$=-1V (red line, bottom axis) and $R_{xx}(V_{mg})$ at $V_{bg}$=-40V. Inset: optical image of a device. (d). Hall resistance vs $V_{bg}$ at $B$=30T, showing well-quantized QH plateaus at filling factors -1, -2, -3 and -4.

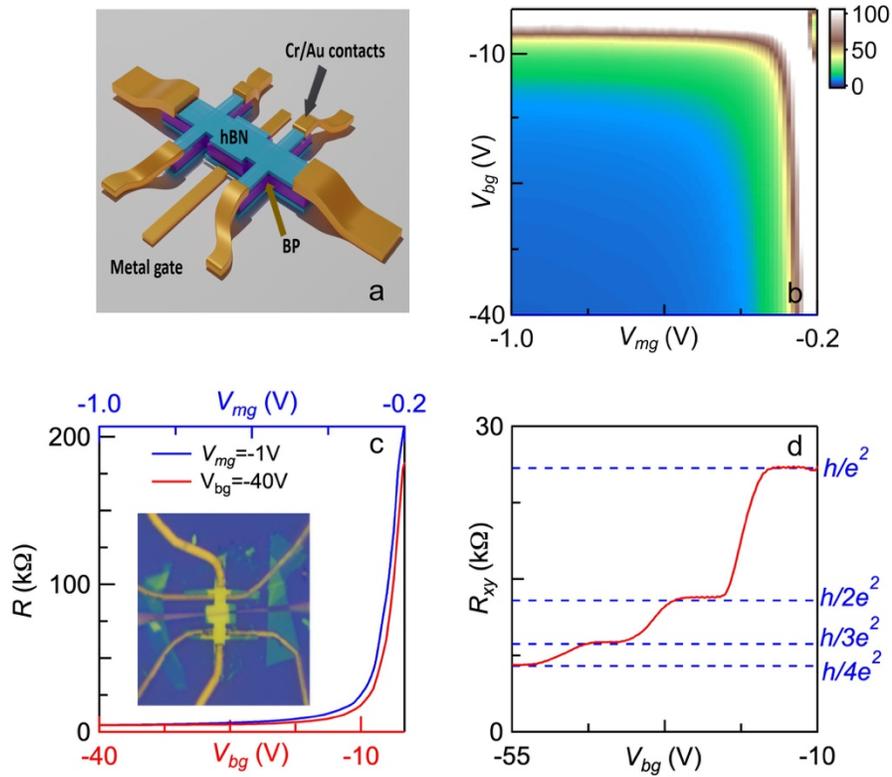

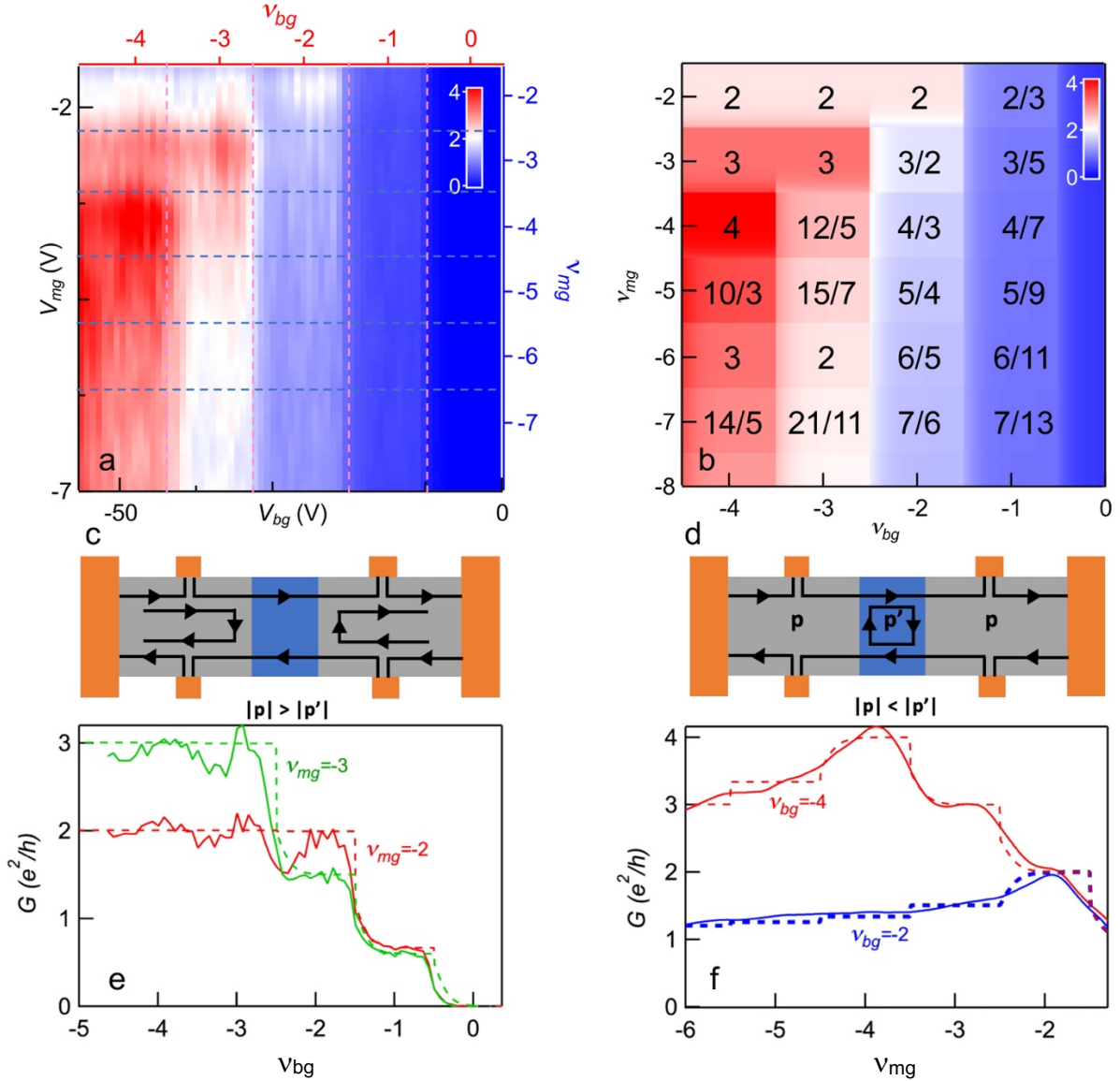

Fig. 2. (a) Conduction $G$ of device D1 (in unit of $e^2/h$) versus back gate voltage $V_{bg}$ and metal gate voltage $V_{mg}$ at $B=30$T, the filling factors of metal-gated region $\nu_{mg}$ and of the back-gated regions $\nu_{bg}$ are labeled on the right and top axes, respectively. (b) Calculated $G(\nu_{bg}, \nu_{mg})$ using Eq. (1), where the number within each square corresponds to expected conductance plateau values. (c-d) Schematic of edge state transport when $|\nu_{mg}|<|\nu_{bg}|$ and $|\nu_{mg}|>|\nu_{bg}|$, respectively. (e). Experimental data $G(\nu_{bg})$ at $\nu_{mg}=-2$ and $-3$ (solid lines) and corresponding theoretical curves (dashed lines) (f) $G(\nu_{mg})$ at $\nu_{bg}=-2$ and $\nu_{bg}=-4$ (solid lines) and corresponding theoretical curves (dashed lines).

Fig. 3. (a) Two-terminal conductance $G$ of device D2 (in unit of $e^2/h$) versus back gate voltage $V_{bg}$ and metal gate voltage $V_{mg}$ at $B=30$T. The filling factors of metal-gated region $\nu_{mg}$ and of the back-gated regions $\nu_{bg}$ are labeled on the left and top axes, respectively. The dashed lines are drawn to approximately delineate regions with different filling factors. (b) Calculated $G(\nu_{bg}, \nu_{mg})$ using full equilibration (left panel) and partial equilibration model (right panel), respectively. (c) Experimental data $G(\nu_{mg})$ (red solid line), theoretical curve using full equilibration model (blue dotted line) and theoretical curve using partial equilibration model (red dashed line) at $\nu_{bg} = -1$. (d) Similar to (c), except at $\nu_{bg} = -2$. Inset: Schematic of edge states with spin polarizations.

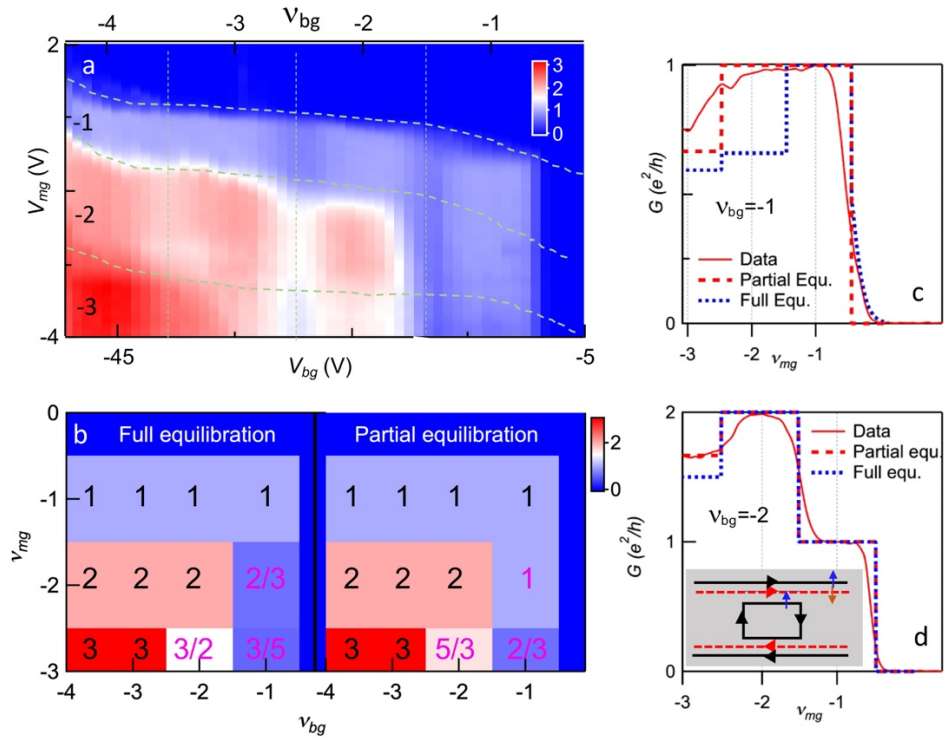

Fig. 4. (a) Calculated $R_{xx}(v_{bg}, v_{mg})$ map using Eq. (3). (b) Longitudinal resistance $R_{xx}(V_{mg}, V_{bg})$ of device D2 at $B=30$T. (c) Experimental data $R_{xx}(v_{mg})$ (red solid line), theoretical curve using full equilibration model (blue dashed line) and theoretical curve using partial equilibration model (red dotted line) at $v_{bg} = -1$. (d) Similar to (c), except at $v_{bg} = -2$.

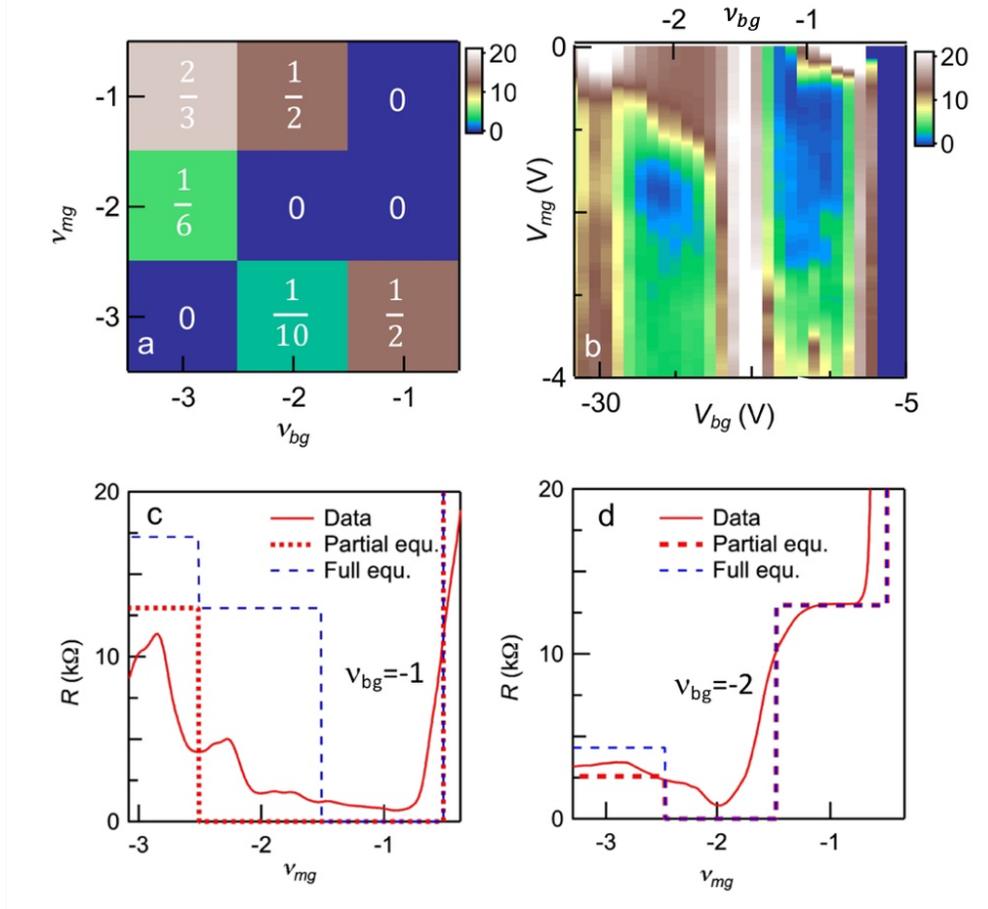

Fig. 5. (a) $G(v_{mg})$ at $v_{bg}=-4$ for device D1. (b) $G(v_{mg})$ at $v_{bg}=-3$ for device D2.

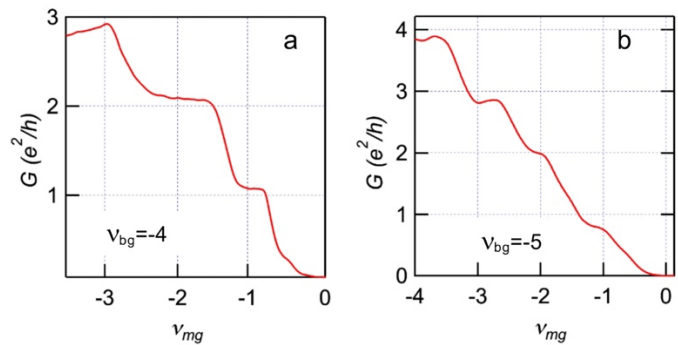